\documentclass[aps,prd,floatfix,nofootinbib,preprintnumbers]{revtex4}

\usepackage[english]{babel}
\usepackage{amsmath,amssymb}
\usepackage[dvips]{graphicx}
\usepackage{rotating}

\def\rpv{$R_p \hspace{-1em}/\;\:\hspace{0.2em}$}

\def\lsim{\raise0.3ex\hbox{$\;<$\kern-0.75em\raise-1.1ex\hbox{$\sim\;$}}}

\newcommand{\AddrAHEP}{%
  AHEP Group, Institut de F\'{\i}sica Corpuscular --
  C.S.I.C. \& Universitat de Val{\`e}ncia \\
  Edificio Institutos de Paterna, Apt 22085, E--46071 Valencia, Spain}
\newcommand{\AddrWur}{%
Institut f\"ur Theoretische Physik und Astronomie, 
Universit\"at W\"urzburg\\
Am Hubland, 
97074 Wuerzburg}

\begin{document}

\preprint{IFIC/09-06}
\title{Majoron emission in muon and tau decays revisited}

\author{M.~Hirsch} \email{mahirsch@ific.uv.es} \affiliation{\AddrAHEP}

\author{J. Meyer} \email{jochen.meyer@physik.uni-wuerzburg.de}
\affiliation{\AddrWur}

\author{W.\ Porod} \email{porod@physik.uni-wuerzburg.de}
\affiliation{\AddrWur}

\author{A.~Vicente} \email{Avelino.Vicente@ific.uv.es} \affiliation{\AddrAHEP}


\begin{abstract}
\vspace*{1cm}
In models where the breaking of lepton number is spontaneous a massless 
Goldstone boson, the Majoron ($J$), appears. We calculate the theoretically 
allowed range for the branching ratios of Majoron emitting charged lepton 
decays, such as Br($\mu \to e J$) and Br($\mu \to e J \gamma$), in a 
supersymmetric model with spontaneous breaking of R-parity. Br($\mu\to eJ$) 
is maximal in the same region of parameter space for which the lightest 
neutralino decays mainly invisibly. A measurement of Br($\mu\to eJ$) thus 
potentially provides information on R-parity violation complementary to 
accelerator searches. We also briefly discuss existing bounds and prospects 
for future improvements on the Majoron coupling to charged leptons. 

\end{abstract}

\maketitle

\section{Introduction}

Spontaneous breaking of lepton number leads to a massless Goldstone boson, 
the Majoron ($J$) \cite{Chikashige:1980ui,Gelmini:1980re,Aulakh:1982yn}. 
There are two well-known experimental probes for the Majoron: The first 
is the invisible width of the $Z^0$ boson, very precisely measured at LEP 
\cite{Amsler:2008zz}. The second is neutrinoless double beta decay 
\cite{Georgi:1981pg}. The NEMO-3 collaboration, for example, has published 
limits on half-lives for Majoron-emitting neutrinoless double beta 
decay for a number of isotopes \cite{Arnold:2006sd}. In addition, 
there are different astrophysical constraints on the Majoron from 
the cooling of red giant stars and supernovae 
\cite{Raffelt:1990yz,Kachelriess:2000qc}. 

Another interesting possibility to search for Majorons, namely charged 
lepton decays with Majoron emission, has attracted considerably less 
attention. Indeed, the limits on $l_i\to l_j J$ quoted by the Particle 
Data Group \cite{Amsler:2008zz} are all based on experimental data which 
is now more than 20 years old. Probably this apparent lack of interest 
from the experimental side is due to the fact that both, the triplet 
\cite{Gelmini:1980re} and the doublet Majoron \cite{Aulakh:1982yn}, are 
ruled out by LEP data, while the (classical) singlet Majoron model 
\cite{Chikashige:1980ui} predicts Majoron-neutrino and Majoron-charged-lepton 
couplings which are unmeasurably small. 

Supersymmetry (SUSY) is a theoretically well motivated extension of the 
standard model. With the first data taking of the LHC only months 
away, searches for SUSY will gain momentum soon. One of the many virtues 
of SUSY is the fact that the minimal supersymmetric extension of the 
standard model (MSSM) provides an interesting candidate for the cold 
dark matter (CDM), usually assumed to be the lightest neutralino ($\chi^0_1$) 
{\em if R-parity is conserved.} A stable, electrically 
neutral lightest supersymmetric particle (LSP) will escape the detector 
and lead to the famous missing momentum signal, upon which standard 
SUSY searches are based. If R-parity is broken, the LSP decays and 
the CDM candidate is lost. For explicit R-parity violation (\rpv), the 
missing energy signal is degraded, but a larger number of jets and 
charged lepton final states should make discovery of \rpv a (comparatively) 
easy task. In {\em spontaneous} \rpv (s-\rpv), however, the lightest 
neutralino can decay invisibly through $\chi^0_1 \to J \nu$. As pointed 
out in \cite{Hirsch:2006di,Hirsch:2008ur}, if the scale of \rpv is low, 
this decay mode can be easily dominant and s-\rpv can be confused 
with a standard MSSM with $R_p$ conserved. 

Here, we revisit $l_i\to l_j J$ within spontaneous R-parity violation. 
Our calculation is based on the model of \cite{Masiero:1990uj}. In 
this model the Majoron is mainly singlet, thus escaping the LEP bounds. 
This is different from the original spontaneous model \cite{Aulakh:1982yn}, 
which used the left-sneutrinos  to break R-parity. Nevertheless, in our 
model the Majoron can play an important role phenomenologically. In 
\cite{Romao:1991tp} $l_i\to l_j J$ was calculated for a tau neutrino 
mass of $m_{\nu_\tau} \simeq $ MeV. Here we show that (a) despite 
the fact that current neutrino mass bounds are of the order of eV 
or less, theoretically $\mu\to e J$ can be (nearly) arbitrarily large 
in s-\rpv, and (b) $\mu\to e J$ is large in the same part of SUSY 
parameter space where the invisible neutralino decay is large, 
making the discovery of R-parity violation at the LHC difficult. 
Br($\mu\to e J$) thus gives complementary information to accelerator 
experiments.

At the same time, the MEG experiment \cite{meg} has started taking 
data. MEG is optimised to search for Br($\mu\to e \gamma$) with a 
sensitivity of Br($\mu\to e \gamma$) $\sim$ (few) $10^{-14}$. 
While the impressive statistics of the experiment should allow, 
in principle, to improve the existing bound on Br($\mu\to e J$) 
\cite{Amsler:2008zz} by a considerable margin, the experimental 
triggers and cuts make it necessary to resort to a search for the 
radiative Majoron emission mode, Br($\mu\to e J \gamma$), if one 
wants to limit (or measure) the Majoron-charged-lepton coupling. 
We therefore also calculate Br($\mu\to e J \gamma$).

In the next section, we will briefly discuss s-\rpv and give 
an approximative, analytical estimation of Br($l_i\to l_j J$) 
and Br($l_i\to l_j J \gamma$). We then present our numerical 
results in section (\ref{sec:num}), showing the correlation 
of Br($l_i\to l_j J$) with the invisible branching ratio of 
the lightest neutralino decay. In section (\ref{sec:exp}) we then 
turn to a brief discussion of existing experimental bounds and 
comment on how a limit on Br($\mu\to e J \gamma$) might be used to 
put a bound on the Majoron-charged-lepton coupling. We then 
close with a brief conclusion in section (\ref{sec:cncl}).

\section{Spontaneous R-parity breaking}
\label{sec:th}

The model we consider \cite{Masiero:1990uj} extends the particle 
spectrum of the MSSM by three additional singlet superfields, 
$\widehat\nu^c$, $\widehat S$ and $\widehat\Phi$, with lepton 
number assignments of $L=-1,1,0$ respectively. The superpotential 
can be written as 
\begin{eqnarray} %
{\cal W} &=& h_U^{ij}\widehat Q_i \widehat U_j\widehat H_u
          +  h_D^{ij}\widehat Q_i\widehat D_j\widehat H_d
          +  h_E^{ij}\widehat L_i\widehat E_j\widehat H_d \nonumber
\\
        & + & h_{\nu}^{i}\widehat L_i\widehat \nu^c\widehat H_u
          - h_0 \widehat H_d \widehat H_u \widehat\Phi
          + h \widehat\Phi \widehat\nu^c\widehat S +
          \frac{\lambda}{3!} \widehat\Phi^3 .
\label{eq:Wsuppot}
\end{eqnarray}
Strictly speaking only 
$\widehat\nu^c$ is necessary to spontaneously break $R_p$. The inclusion 
of $\widehat S$ and $\widehat\Phi$ allows to construct a superpotential 
which purely consists of trilinear terms, thus potentially solving also 
the $\mu$ problem of the MSSM. At low energy various fields acquire vacuum 
expectation values (vevs). Besides the usual MSSM Higgs 
boson vevs $v_d$ and $v_u$, these are 
$\langle \Phi \rangle = v_{\phi}/\sqrt{2}$, 
$\langle {\tilde \nu}^c \rangle = v_R/\sqrt{2}$,
$\langle {\tilde S} \rangle = v_S/\sqrt{2}$ and 
$\langle {\tilde \nu}_i \rangle = v_{L_i}/\sqrt{2}$. Note, that $v_R \ne 0$ 
generates effective bilinear terms $\epsilon_i = h_{\nu}^i v_R/\sqrt{2}$ 
and that $v_R$, $v_S$ and $v_{L_i}$ violate lepton number as well as 
R-parity. The observed smallness of neutrino masses guarantees that the 
\rpv operators generated by these vevs are small. The smallness of 
$\epsilon_i$ implies that either $h_{\nu}^i$ or $v_R$ is small, 
but not necessarily both. 

Details of the model, such as mass matrices and couplings, can be 
found in \cite{Hirsch:2004rw,Hirsch:2005wd}, for the phenomenology 
of the LSP decay in this model see \cite{Hirsch:2006di,Hirsch:2008ur}. 
For brevity in the following we will concentrate on only a few, relevant 
aspects of the phenomenology of this model: The neutrino mass matrix, 
the lightest neutralino decay to Majorons and charged lepton decays. 

The main motivation to study R-parity breaking supersymmetry certainly 
is that \rpv generates neutrino masses and thus contains a possible 
explanation for the observed neutrino oscillation data. For the 
spontaneous model defined in eq. (\ref{eq:Wsuppot}), the effective 
neutrino mass matrix at tree-level can be cast into a very simple 
form 
\begin{equation}
 -(\boldsymbol{m_{\nu\nu}^{\rm eff}})_{ij} = a \Lambda_i \Lambda_j + 
     b (\epsilon_i \Lambda_j + \epsilon_j \Lambda_i) +
     c \epsilon_i \epsilon_j\,.
\label{eq:eff}
\end{equation}
Here, $\Lambda_i = \epsilon_i v_d + v_{L_i} \mu$, with 
$\mu=h_0v_{\phi}/\sqrt{2}$. The coefficients 
$a$, $b$ and $c$ are defined as
\begin{eqnarray}\label{def:abc}
a=\frac{m_\gamma h^2 v_\phi}{4 \sqrt{2} Det(M_H)} 
(-h v_R v_S+\frac{1}{2} \lambda v_\phi^2+h_0 v_d v_u), \\ \nonumber
b=\frac{m_\gamma h^2 \mu}{4 Det(M_H)} v_u (v_u^2-v_d^2), \\ \nonumber 
c=\frac{h^2 \mu}{Det(M_H)} v_u^2 (2 M_1 M_2 \mu - m_\gamma v_d v_u).
\end{eqnarray}
$Det(M_H)$ is the determinant of the ($7,7$) matrix of the heavy 
neutral states (the four MSSM states, $\tilde B$, $\tilde W$ and 
$\tilde H_{u,d}$, plus the three fermionic components of the new 
singlet superfields of eq. (\ref{eq:Wsuppot})) 
\begin{equation}\label{def:detmh}
Det(M_H)=\frac{1}{16} h_0 h^2  v_\phi^2 \big[ 4(2 M_1 M_2 \mu 
- m_\gamma v_d v_u)(-h v_R v_S+\frac{1}{2} 
\lambda v_\phi^2+h_0 v_d v_u)-h_0 m_\gamma (v_u^2-v_d^2)^2 \big]
\end{equation}
and $v^2=v_u^2+v_d^2$. The ``photino'' mass parameter is defined 
as $m_{\gamma} = g^2M_1 +g'^2 M_2$. Since $\mu=h_0v_{\phi}/\sqrt{2}$ 
$Det(M_H) \propto v_{\phi}^3$ in the limit of large $v_{\phi}$. 
One can easily fit the observed neutrino masses and angles using 
eq. (\ref{eq:eff}), see \cite{Hirsch:2008ur} and the short discussion 
in the next section.

From a phenomenological point of view the most important difference 
between spontaneous and explicit R-parity violating models is the 
appearance of the Majoron. The pseudo-scalar sector of the model we 
consider has eight different eigenstates. Two of them are Goldstone 
bosons. The standard one is eaten by the $Z^0$ boson, the remaining 
state is identified with the Majoron. In the limit $v_{L_i} \ll v_R,v_S$ 
the Majoron profile is given by the simple expression 
\begin{equation}\label{smplstmaj}
R_{Jm}^{P^0} \simeq 
\big(0,0,\frac{v_{L_k}}{V},0,\frac{v_S}{V},-\frac{v_R}{V}\big). 
\end{equation}
Here, $V=\sqrt{v_R^2+v_S^2}$ and terms of order $\frac{v_L^2}{V v}$, 
where $v_L^2 = \sum_i v_{L_i}^2$, have been neglected. 

Majorons are weakly coupled, thus potentially lead to a decay mode 
for the lightest neutralino which is invisible. Neutralino-Majoron 
couplings can be calculated from the general coupling 
${\chi}_i^0-{\chi}_j^0-P^0_k$
\begin{equation}\label{defnnp}
\mathcal{L}=\frac{1}{2} \bar{\chi}_i^0 
\big( O_{Lij}^{nnp} P_L + O_{Rij}^{nnp} P_R \big) \chi_j^0 P_k^0.
\end{equation}
Mixing between the neutralinos and the neutrinos then leads to a 
coupling $\chi^0_1-\nu_k-J$. In the limit $v_R, v_S \gg \epsilon_i, v_{L_i}$ 
one can derive a very simple approximation formula for 
$O_{\tilde\chi^0_1\nu_kJ}$. It s given by
\cite{Hirsch:2008ur}
\begin{equation}
\label{eq:majcl}
|O_{\tilde\chi^0_1\nu_kJ}| \simeq  - \frac{{\tilde \epsilon}_k}{V}N_{14} +
\frac{{\tilde v}_{L_k}}{2 V}(g' N_{11} - g N_{12})
+ \cdots, 
\end{equation}
where the dots stand for higher order terms neglected here and $N$ is 
the matrix which diagonalizes the (MSSM) neutralino mass matrix. 
${\tilde\epsilon} = U_{\nu}^T\cdot{\vec\epsilon}$ and 
${\tilde v}_{L}=U_{\nu}^T\cdot{\vec v_L}$. Here $(U_{\nu})^T$ is the matrix 
which diagonalizes either the part of the ($3,3$) effective neutrino mass 
matrix, proportional to $a$ or $c$, depending on which gives the larger 
eigenvalue. 
Eq. (\ref{eq:majcl}) shows that for constant ${\tilde \epsilon}$ 
and ${\tilde v}_{L}$, $O_{\tilde\chi^0_1\nu_kJ} \rightarrow 0$ as $v_R$ 
goes to infinity. This is as expected, since for $v_R\rightarrow\infty$ 
the spontaneous model approaches the explicit bilinear model. We note 
that, in addition to the Majoron there is also a rather light singlet 
scalar, called the ``scalar partner'' of the Majoron in \cite{Hirsch:2005wd}, 
$S_J$. The lightest neutralino has a coupling $O_{\tilde\chi^0_1\nu_kS_J}$, 
which is of the same order as $O_{\tilde\chi^0_1\nu_kJ}$. Since $S_J$ 
decays to nearly 100 $\%$ to two Majorons, this decay mode contributes 
sizeably to the invisble width of the lightest neutralino, for more 
details see \cite{Hirsch:2008ur}. 

The decays $l_i\to l_j J$ can be calculated from the general 
coupling $\chi^+_i-\chi^-_j-P^0_k$. In the limit of small 
R-parity violating parameters the relevant interaction lagrangian 
for the $l_i - J - l_j$ coupling is given by
\begin{equation}
\mathcal{L}= \bar{l}_i 
\big( O_{LijJ}^{ccp} P_L + O_{RijJ}^{ccp} P_R \big) l_j J
\end{equation}
with
\begin{eqnarray}\label{cpl_llJ}
O_{RijJ}^{ccp} &=& - \frac{i (h_E)^{jj}}{\sqrt{2} V} \big[ 
                    \frac{v_d v_L^2}{v^2} \delta_{ij} 
  + \frac{1}{\mu^2}(C_1 \Lambda_i \Lambda_j + C_2 \epsilon_i \epsilon_j 
+ C_3 \Lambda_i \epsilon_j + C_4 \epsilon_i \Lambda_j) \big]\nonumber \\
O_{LijJ}^{ccp} &=& \big(O_{RjiJ}^{ccp}\big)^*.
\end{eqnarray}
The $C$ coefficients are different combinations of MSSM parameters
\begin{eqnarray}
C_1 &=& \frac{g^2}{2 Det_+^2}
(- g^2 v_d v_u^2 - v_d \mu^2 + v_u M_2 \mu) \\
C_2 &=& -2 v_d \hskip10mm
C_3 = - \frac{g^2 v_d v_u}{Det_+} \hskip10mm
C_4 = 1 - \frac{g^2 v_d v_u}{2 Det_+} \nonumber
\end{eqnarray}
where $Det_+$ is the determinant of the MSSM chargino mass matrix
$Det_+ = M_2 \mu - \frac{1}{2} g^2 v_d v_u$. Eq. (\ref{cpl_llJ}) 
shows that one expects large partial widths to Majorons, if 
$v_R$ is low. 

For a charged lepton $l_i$, with polarization vector $\vec P_i$, the decay
$l_i \rightarrow l_j J$ has a differential decay width given by
\begin{eqnarray}\label{decwid}
\frac{d\Gamma (l_i \rightarrow l_j J)}{d \cos \theta} &=& 
             \frac{m_i^2 - m_j^2}{64 \pi m_i^3} 
   \big[ |O_{LijJ}^{ccp}|^2 \big( m_i^2 + m_j^2 \pm (m_i^2 - m_j^2) 
           P_i \cos \theta \big) \nonumber \\
&& + |O_{RijJ}^{ccp}|^2 \big( m_i^2 + m_j^2 \mp (m_i^2 - m_j^2) 
        P_i \cos \theta \big) \\
&& + 4 m_i m_j Re( {O_{LijJ}^{ccp}}^* O_{RijJ}^{ccp} ) \big] \nonumber
\label{eq:dgampol}
\end{eqnarray}
where $\theta$ is the angle between the polarization vector $\vec P_i$ 
and the momentum $\vec p_j$ of the charged lepton in the final state, 
and $P_i = |\vec P_i|$ is the polarization degree of the decaying 
charged lepton.

\noindent
In the limit $m_j \simeq 0$ the expression \eqref{decwid} simplifies to
\begin{equation}
\frac{d\Gamma (l_i \rightarrow l_j J)}{d \cos \theta} 
= \frac{m_i}{64 \pi} |O_{LijJ}^{ccp}|^2 \big( 1 \pm P_i \cos \theta \big)
\end{equation}
since $|O_{RijJ}^{ccp}|^2 \propto ( h_E^{jj} )^2 \propto m_j^2$. The 
angular distribution of the Majoron emitting lepton decay is thus very 
similar to the standard model muon decay \cite{Amsler:2008zz}, 
up to corrections of the order $(m_j/m_i)^2$, which are negligible in 
practice. 

\begin{figure}[htbp]
\begin{center}
\vspace{5mm}
\includegraphics[width=50mm,height=30mm]{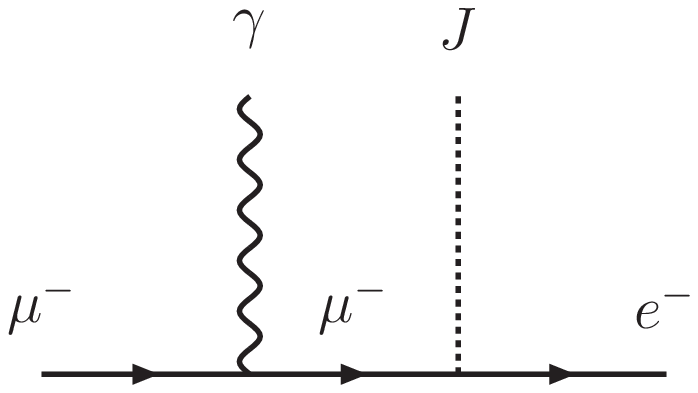}
\hskip15mm
\includegraphics[width=50mm,height=30mm]{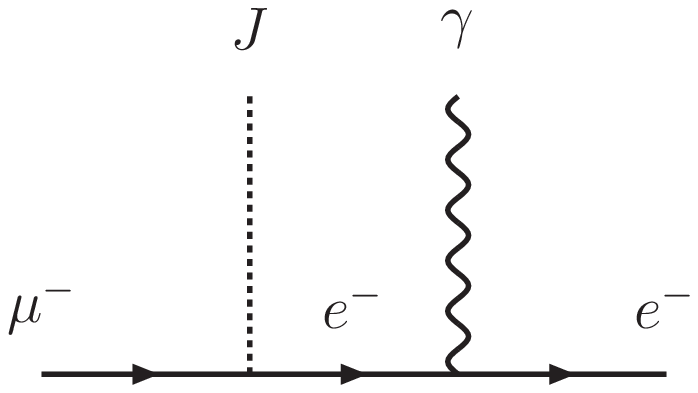}
\end{center}

\caption{Feynman diagrams for the decay $\mu \to e J \gamma$. 
As in the standard model radiative decay $\mu \to e {\bar\nu}\nu\gamma$ 
these diagrams contain an infrared divergence for $m_{\gamma}=0$, see 
text.}

\label{fig:feynG}
\end{figure}

We next consider the decay $\mu \to e J \gamma$. \footnote{Formulas 
for the radiative Majoron decays of the $\tau$ can be found from 
straightforward replacements.} It is induced by the Feyman diagrams 
shown in fig. (\ref{fig:feynG}). 

In the approximation $m_e \simeq 0$ the partial decay width for the 
process $\mu \to e J \gamma$ can be written as
\begin{equation}
\Gamma(\mu \to e J \gamma) = \frac{\alpha }{64 \pi^2} 
              |O_{L\mu eJ}^{ccp}|^2 m_\mu {\cal I}(x_{min},y_{min})
\end{equation}
where ${\cal I}(x_{min},y_{min})$ is a phase space integral given by
\begin{equation}
{\cal I}(x_{min},y_{min}) = \int dx dy f(x,y) = 
\int dx dy \frac{(x-1)(2-xy-y)}{y^2(1-x-y)},
\end{equation}
the dimensionless parameters $x$, $y$ are defined as usual
\begin{equation}
x = \frac{2 E_e}{m_\mu} \quad , \quad y = \frac{2 E_\gamma}{m_\mu}
\end{equation}
and $x_{min}$ and $y_{min}$ are the minimal electron and photon 
energies measured in a given experiment. 

Note that the integral ${\cal I}(x_{min},y_{min})$ diverges for $y_{min}=0$. 
This infrared divergence is well-known from the standard model 
radiative decay $\mu \to e {\bar\nu}\nu\gamma$, and can be taken 
care off in the standard way by introducing a non-zero photon 
mass $m_{\gamma}$. Note that in the limit $m_e=0$ there also appears 
a colinear divergence, just as in the SM radiative decay. Since in any 
practical experiment there is a minimum measurable photon energy, 
$y_{min}$, as well as a minimum measurable photon-electron angle 
($\theta_{e\gamma}$), neither divergence affects us in practice. 
We simply integrate 
from the minimum value of $y$ up to $y_{max}$ when estimating the 
experimental sensitiviy of Br($\mu \to e J \gamma$) on the Majoron 
coupling.

In the calculation of the integral ${\cal I}(x_{min},y_{min})$ one 
has to take into account not only the experimental cuts applied 
to the variables $x$ and $y$, but also the experimental cut for 
the angle between the directions of electron and photon. This 
angle is fixed for kinematical reasons to
\begin{equation}
\cos \theta_{e \gamma} = 1 + \frac{2-2(x+y)}{xy}.
\label{eq:ctheta}
\end{equation}
This relation restricts $x_{max}$ to be $x_{max}\le 1$ as a 
function of $y$ (and vice versa) and to $x_{max}< 1$ for 
$\cos \theta_{e \gamma} > -1$. 

Using the formula for $\Gamma(\mu \to e J)$, in the approximation 
$m_e \simeq 0$,
\begin{equation}
\Gamma(\mu \to e J) = \frac{m_\mu}{32 \pi} |O_{L\mu eJ}^{ccp}|^2
\end{equation}
one finds a very simple relation between the two branching ratios
\begin{equation}
Br(\mu \to e J \gamma) = \frac{\alpha}{2 \pi} {\cal I}(x_{min},y_{min}) 
                          Br(\mu \to e J).
\label{eq:relBr}
\end{equation}
We will use eq. (\ref{eq:relBr}) in section (\ref{sec:exp}) when 
we discuss the relative merits of the two different measurements.

\section{Numerical results}
\label{sec:num}

All numerical results shown in this section have been obtained using the 
program package SPheno \cite{Porod:2003um}, extended to include the new 
singlet superfields $\widehat\nu^c$, $\widehat S$ and $\widehat\Phi$. 
We always choose the \rpv parameters in such a way that solar and 
atmospheric neutrino data \cite{Maltoni:2004ei} are fitted correctly. 
The numerical procedure to fit neutrino masses is the 
following.  For any random choice of MSSM parameters, we can reproduce 
the ``correct'' MSSM value of $\mu$ for a random value of $v_{\phi}$, 
by appropriate choice of $h_0$. For any random set of $h$, $\lambda$, 
$v_S$ and $v_R$, we can then calculate those values of $h_{\nu}^i$ 
and $v_{L_i}$, using eq. (\ref{eq:eff}), such that the corresponding 
$\epsilon_i$ and $\Lambda_i$ give correct neutrino masses and mixing 
angles. In the plots shown below we use $\Lambda_i$ for the atmospheric 
scale and $\epsilon_i$ for the solar scale.

\begin{figure}[htbp]
\begin{center}
\vspace{5mm}
\includegraphics[width=80mm,height=60mm]{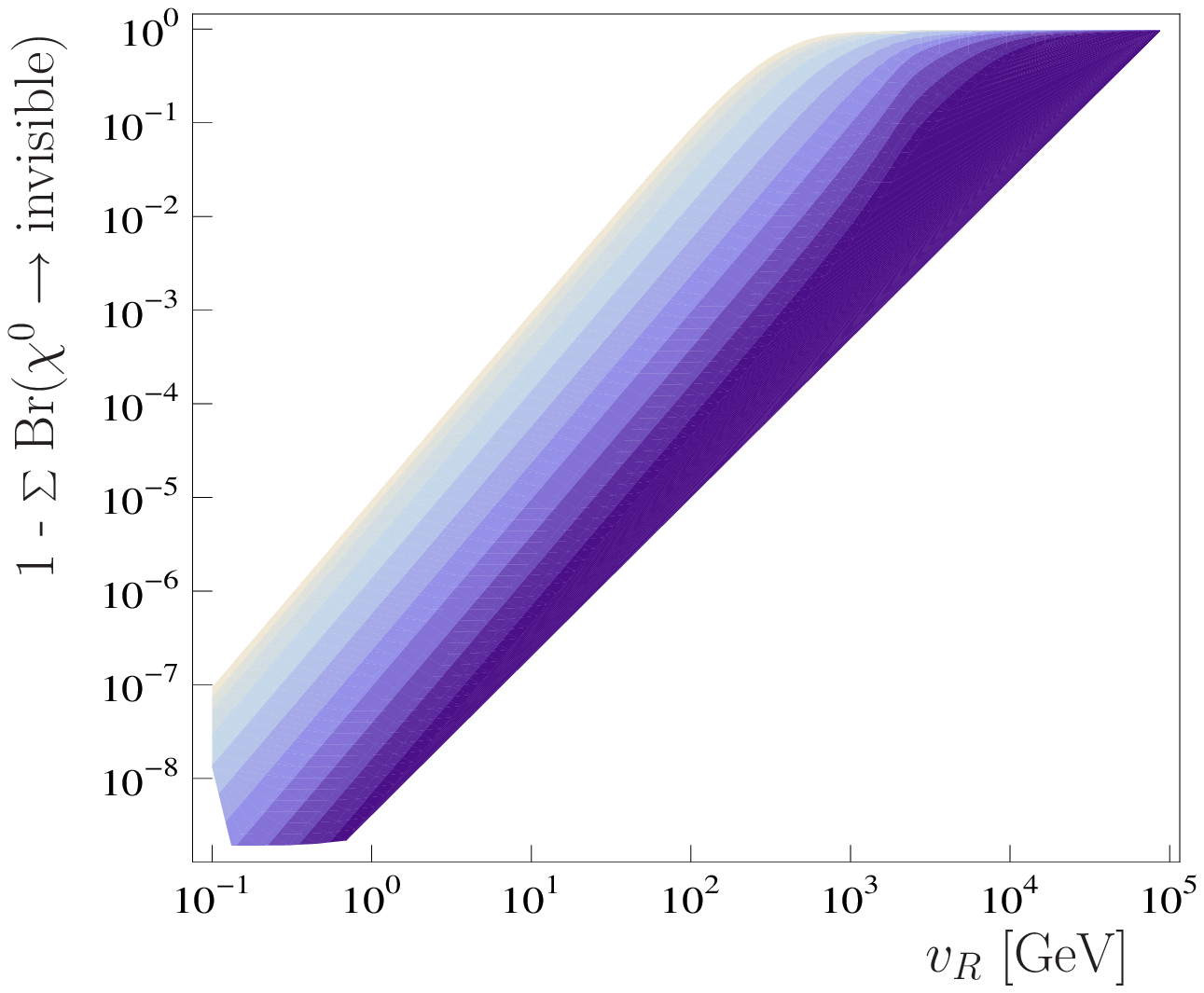}
\includegraphics[width=80mm,height=60mm]{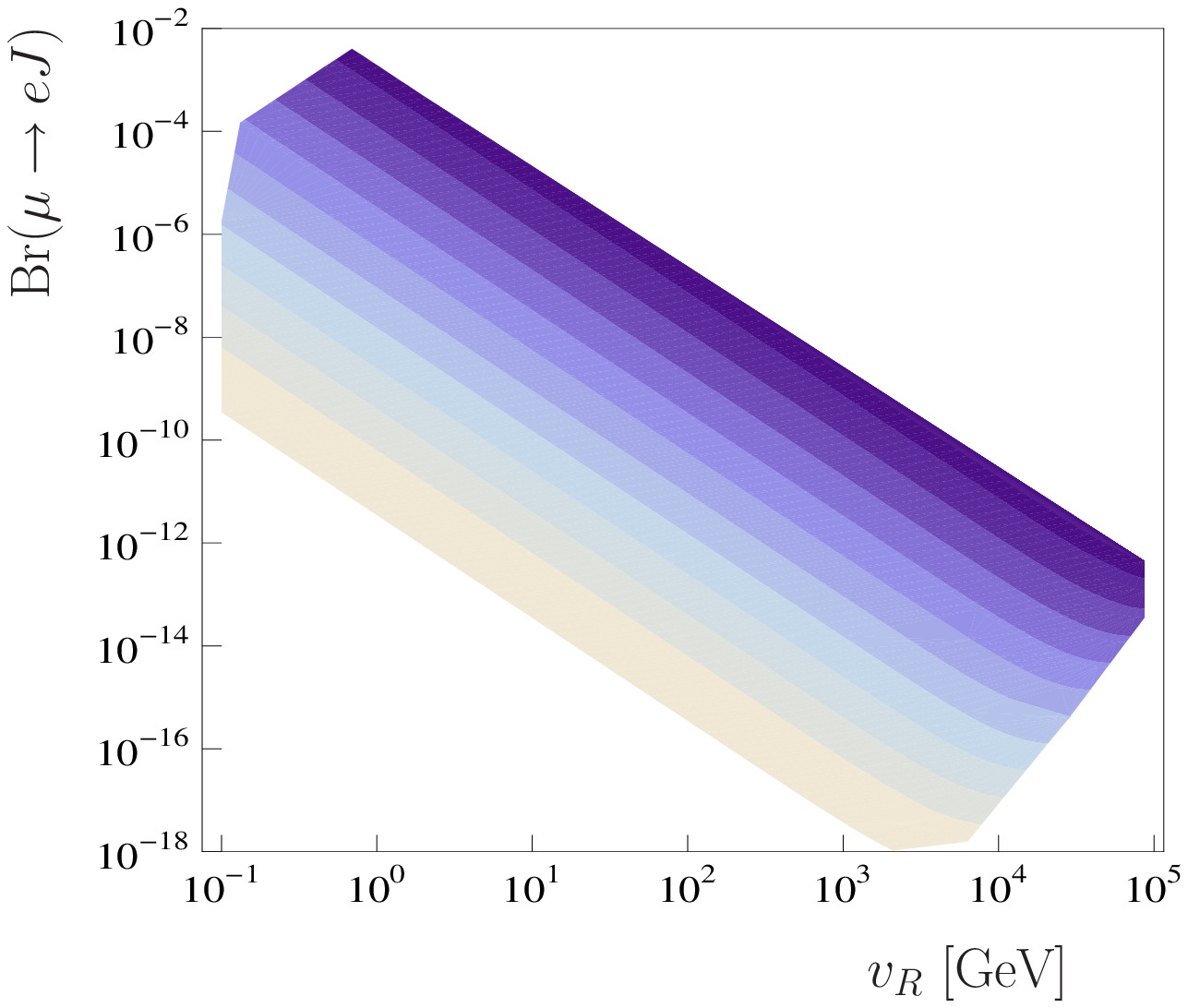}
\end{center}

\caption{Branching ratios for visible lightest neutralino decay (left) 
and branching ratio Br($\mu\to e J$) (right) versus $v_R$ in GeV 
for a number of different choice of $v_{\phi}$ between [$1,10^2$] TeV 
indicated by the different colours. Darker colours indicate larger 
$v_{\phi}$ in a logarithmic scale. mSugra parameters defined in the text. 
There is very little dependence on the actual mSugra parameters, however, 
see discussion and fig. (\ref{fig:VisvMuEj}).}

\label{fig:vis}
\end{figure}

As shown previously \cite{Hirsch:2008ur} if the lightest neutralino is 
mainly a bino, the decay to Majoron plus neutrino is dominant if 
$v_R$ is low. This is demonstrated again for a bino LSP in 
fig. (\ref{fig:vis}), to the left, for a sample point using mSugra 
parameters $m_0=280$~GeV, 
$m_{1/2}=250$~GeV, $\tan\beta=10$, $A_0=-500$~GeV and sgn$(\mu)=+$. 
We stress that this result is independent of the choice of mSugra 
parameters to a large degree \cite{Hirsch:2006di}.  A scan over 
$v_{\phi}$ has been performed in this plot, varying $v_{\phi}$ in 
the huge interval [$1,10^2$] TeV. Large values of $v_{\phi}$ lead 
to small values of the constant $c$ in the neutrino mass matrix, 
see eq. (\ref{eq:eff}). Small $c$ require, for constant neutrino 
masses, large values of $\epsilon_i$, which in turn lead to a 
large invisible width of the neutralino. The largest values of 
$v_{\phi}$ (dark areas) therefore lead to the smallest visible 
neutralino decay branching ratios shown in fig. (\ref{fig:vis}).

Fig. (\ref{fig:vis}), to the right, shows the branching ratio 
Br($\mu\to e J$) as a function of $v_R$ for different values of 
$v_{\phi}$. All parameters have been fixed to the same values as 
shown in the left figure. As the figure demonstrates, small values 
of $v_R$ (and large values of $v_{\phi}$) lead to large values of 
Br($\mu\to e J$). This agrees with the analytic expectation, 
compare to eq. (\ref{cpl_llJ}). 

\begin{figure}[htbp]
\begin{center}
\vspace{5mm}
\includegraphics[width=80mm,height=60mm]{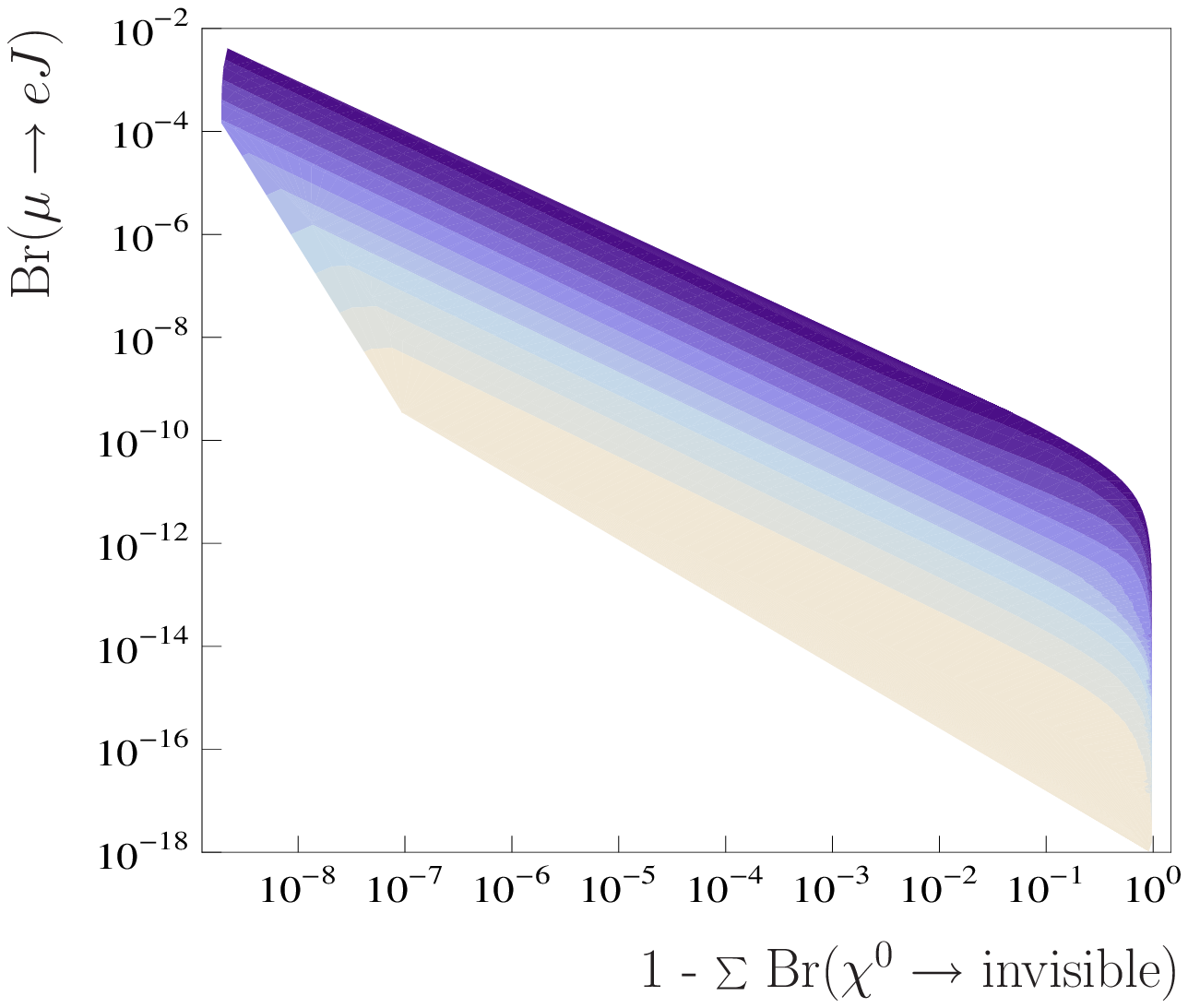}
\includegraphics[width=80mm,height=60mm]{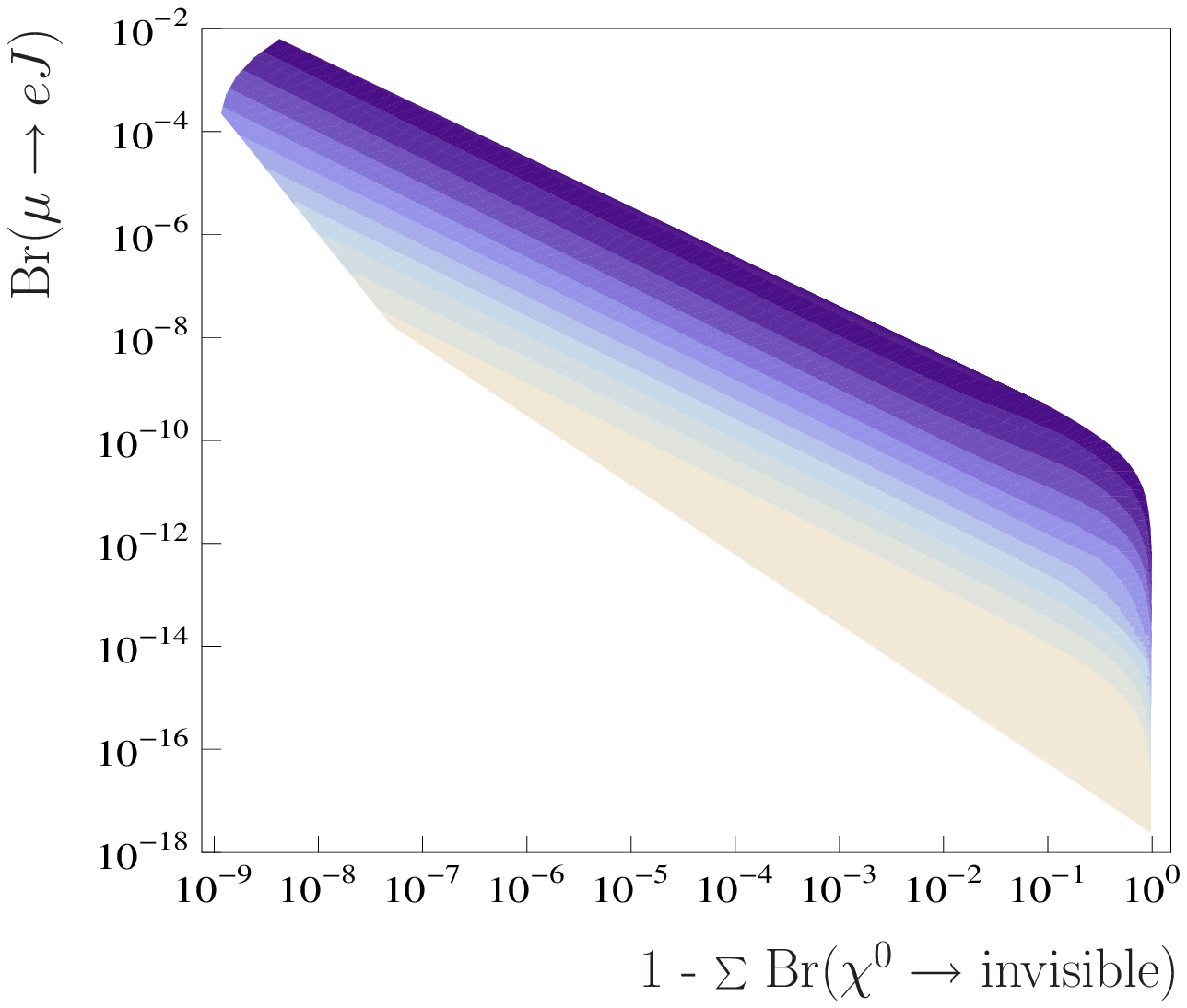}
\end{center}

\caption{Branching ratios for visible lightest neutralino decay versus  
branching ratio Br($\mu\to e J$) for two mSugra points, for various 
choices of $v_{\phi}$, see fig. (\ref{fig:vis}). To the left, same 
mSugra parameters as fig. (\ref{fig:vis}), to the right SPS1a'.}

\label{fig:VisvMuEj}
\end{figure}

Our main result is shown in fig. (\ref{fig:VisvMuEj}). In this 
figure we show Br($\mu\to eJ$) versus the sum of all branching 
ratios of neutralino decays leading to at least one visible particle 
in the final state for two different choices of mSugra parameters.
The similarity of the two plots shows that our result is only 
weakly dependent on the true values of mSugra parameters. We have 
checked this fact also by repeating the calculation for other 
mSugra points, although we do not show plots here.  
As expected Br($\mu\to e J$) anticorrelates with the visible 
bino decay branching fraction and thus probes a complementary 
part in the supersymmetric parameter space. An upper bound on 
Br($\mu\to e J$) will constrain the maximum branching ratio 
for invisible neutralino decay, thus probing the part of parameter 
space where spontaneous R-parity breaking is most easily confused 
with {\em conserved} R-parity at accelerators.

We have checked the points shown in the plots for various 
phenomenological constraints. LEP bounds are trivially fulfilled 
by $v_{L_i} < v_R$. Double beta decay bounds on $g_{\nu\nu J}$ 
\footnote{We will use the symbol 
$g$ when discussing experimental bounds, to differentiate from the 
model dependent couplings $O^{ccp}_L$ and $O^{ccp}_R$ defined in 
section (\ref{sec:th}).} are of 
the order of $10^{-4}$ \cite{Arnold:2006sd} and, since the coupling 
$g_{\nu\nu J}$ is suppressed by two powers of R-parity violating 
parameters, are easily satisfied in our model. More interesting 
is the astrophysical limit on $g_{eeJ}$. 
Ref. \cite{Raffelt:1990yz} quotes a bound of $g_{eeJ} \le 3 \cdot 10^{-13}$. 
Although this bound is derived from the coupling of the Majoron to two 
electrons, thus constraining actually the products $v_{L_e}^2$, 
$\epsilon_e^2$ and $\Lambda_e^2$, whereas Br($\mu\to e J$) is proportional 
to $\epsilon_e\epsilon_{\mu}$ and $\Lambda_e\Lambda_{\mu}$, it still 
leads to a (weak) constraint on Br($\mu\to eJ$), since neutrino physics 
shows that two leptonic mixing angles are large. This requires that either 
$\epsilon_e\sim \epsilon_{\mu}$ {\em or} $\Lambda_e \sim \Lambda_{\mu}$. 
For the case studied in our plots, where $\epsilon_i$ generate the solar 
scale, $\tan^2\theta_{\odot} \simeq 1/2$ requires $\epsilon_e\sim 
\epsilon_{\mu}$. Numerically we then find that $g_{eeJ} \le 3 \cdot 10^{-13}$ 
corresponds to an upper bound on Br($\mu \to e J$) of very roughly 
Br($\mu \to e J$) $\lsim {\rm (few)} \cdot 10^{-5}$. In case the neutrino 
data is fitted with $\vec\epsilon$ for the atmospheric scale, the 
corresponding bound is considerably weaker.

\section{Experimental constraints and Br($\mu\to eJ\gamma$)}
\label{sec:exp}

The Particle Data Group \cite{Amsler:2008zz} cites \cite{Jodidio:1986mz} 
with an upper limit on the branching ratio of Br($\mu\to e X^0$) 
$\le 2.6 \times 10^{-6}$, where $X^0$ is a scalar boson called the 
familon. This constraint does not apply to the Majoron we consider 
here, since it is derived from the decay of polarized muons in a 
direction opposite to the direction of polarization. The authors of 
\cite{Jodidio:1986mz} concentrated on this region, since it minimizes 
events from standard model $\beta$-decay. As shown in eq. (\ref{eq:dgampol}), 
the Majoron emitting decay has a very similar angular distribution 
as the standard model decay, with the signal approaching zero in the 
data sample analyzed by \cite{Jodidio:1986mz}. Nevertheless, from the 
spin processed data shown in fig.(7) of \cite{Jodidio:1986mz}, which 
seems to be in good agreement with the SM prediction, it should 
in principle be possible to extract a limit on Br($\mu\to e J$). 
From this figure we estimate very roughly that this limit should 
be about one order of magnitude less stringent than the one for 
familon decay. For a better estimate a re-analysis of this data, 
including systematic errors, would be necessary.

Ref. \cite{Picciotto:1987pp} searched for Majorons in the 
decay of $\pi\to e \nu J$, deriving a limit of Br($\pi\to e \nu J$) 
$\le 4 \cdot 10^{-6}$. Since the experimental cuts used in this 
paper \cite{Picciotto:1987pp} are designed to reduce the standard 
model background from the decay chain $\pi\to\mu \to e$, the 
contribution from on-shell muons is reduced by about five orders 
of magnitude. The limit then essentially is a limit on the 
Majoron-neutrino-neutrino coupling, $g_{\nu\nu J}$, leaving only a
very weak constraint on the coupling $g_{\mu e J}$. 
Also an analysis searching for Br($\mu\to eJ\gamma$) has been published 
previously \cite{Goldman:1987hy}. From a total data sample of 
$8.15 \cdot 10^{11}$ stopped muons over the live time of the 
experiment \cite{Goldman:1987hy} derived a limit on Br($\mu\to eJ\gamma$) 
of the order of Br($\mu\to eJ\gamma$) $\le 1.3 \cdot 10^{-9}$. 
For the cuts used in this analysis, we calculate ${\cal I} \simeq 10^{-3}$. 
Thus, see eq.(\ref{eq:relBr}), this limit translates into only a 
rather weak bound Br($\mu\to eJ$) $\le 1.1\cdot 10^{-3}$.

The MEG experiment \cite{meg} has a muon stopping rate of $(0.3-1)\cdot 10^8$ 
per second and expects a total of the order of $10^{15}$ muons over 
the expected live time of the experiment. An analysis of {\em electron 
only} events near the endpoint should therefore allow, in principle, 
to improve the existing limits on Br($\mu\to eJ$) by an estimated ($2-3$) 
orders of magnitude, if systematic errors can be kept under control. 
However, the MEG experiment, as it is designed to search for 
Br($\mu\to e\gamma$), uses a trigger that requires a photon in the 
event with a minimum energy of $E_{\gamma}^{min}\ge 45$ MeV. 
MEG data can thus constrain the Majoron-charged-lepton coupling only 
via searching for $\Gamma$($\mu\to eJ\gamma$).

\begin{figure}[htbp]
\begin{center}
\vspace{5mm}
\includegraphics[width=80mm,height=60mm]{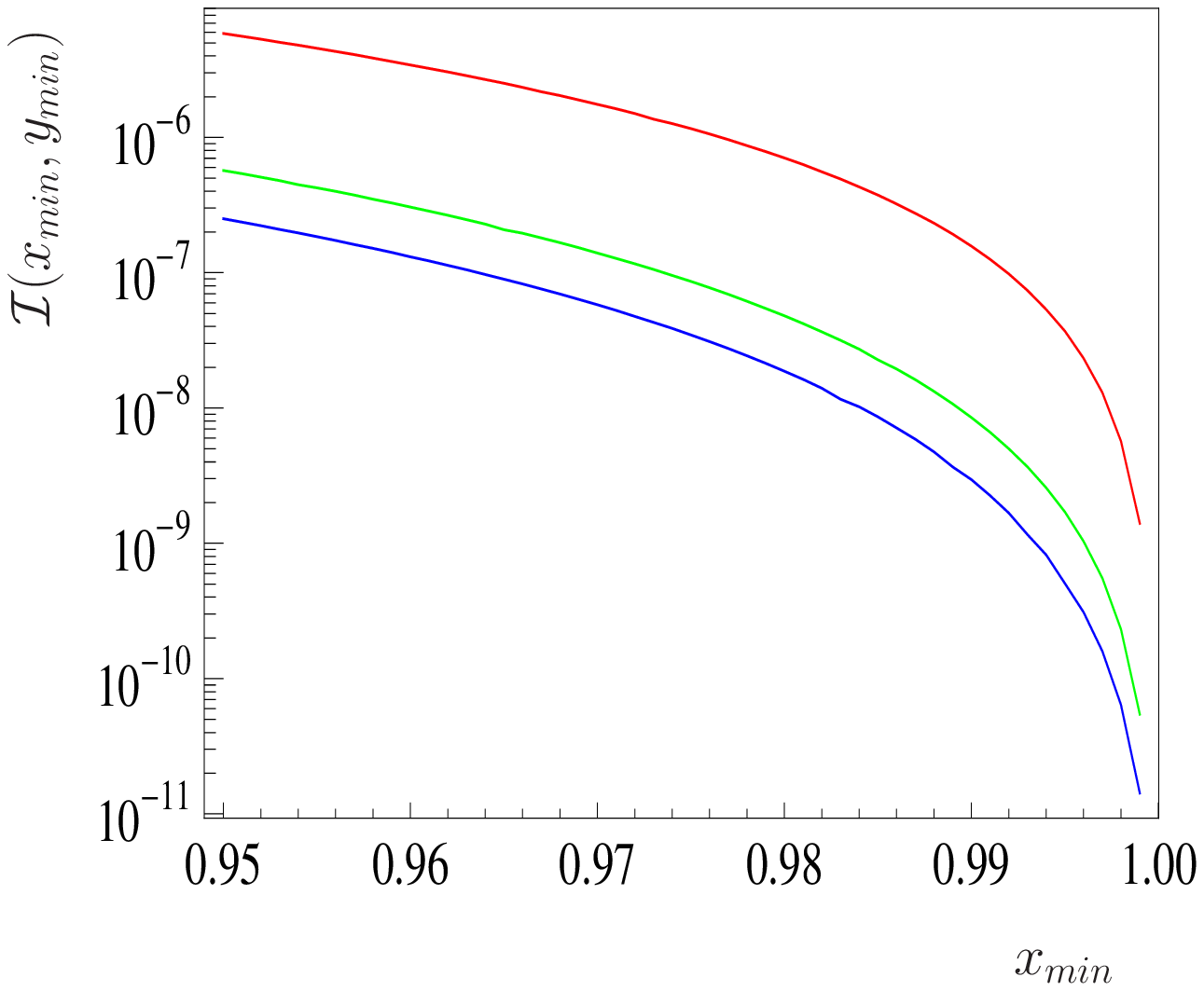}
\includegraphics[width=80mm,height=60mm]{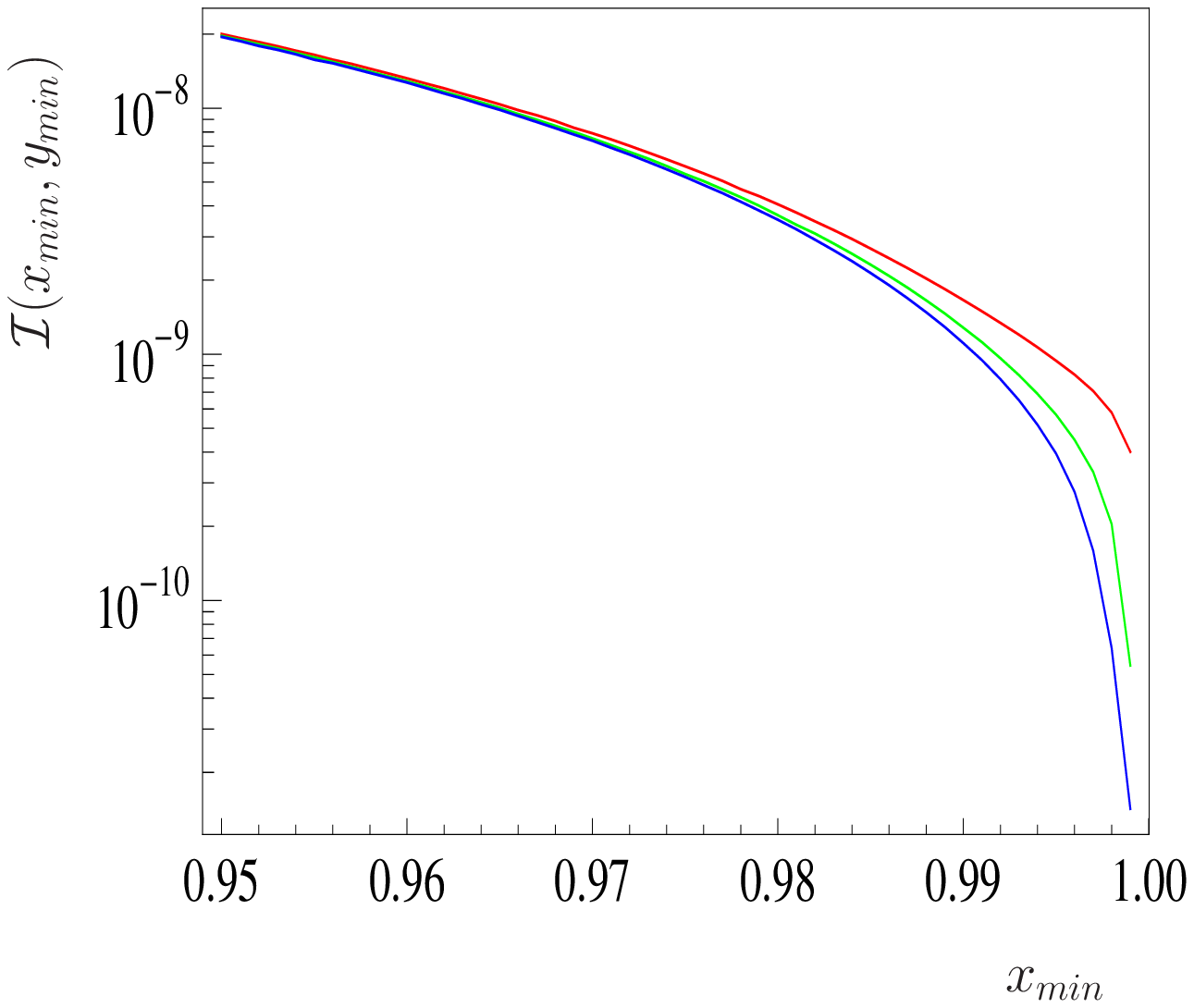}
\end{center}

\caption{The phase space integral for the decay $\mu\to e J \gamma$ 
as a function of $x_{min}$ for three different values of $y_{min}= 
0.95, 0.99, 0.995$ from top to bottom and for two different values of 
$\cos\theta_{e\gamma}$. To the left $\cos\theta_{e\gamma}=-0.99$, 
to the right $\cos\theta_{e\gamma}=-0.99997$.}

\label{fig:int}
\end{figure}

Fig. (\ref{fig:int}) shows the value of the phase space integral 
${\cal I}(x_{min},y_{min})$ as a function of $x_{min}$ for three 
different values of $y_{min}$ and for two choices of $\cos\theta_{e\gamma}$. 
The MEG proposal describes the cuts used in the search for 
$\mu\to e \gamma$ as $x_{min}\ge 0.995$, $y_{min}\ge 0.99 $  
and $|\pi - \theta_{e\gamma}| \le$ 8.4 mrad. For these values we find 
a value of ${\cal I} \simeq 6 \cdot 10^{-10}$. A limit for 
Br($\mu\to e \gamma$) of Br($\mu\to e \gamma$) $\le 10^{-13}$ then 
translates into a limit of Br$(\mu\to e J) \le 0.14$, obviously 
not competitive. To improve upon this bound, it is necessary to 
relax the cuts. For example, relaxing the cut on the opening angle 
to $\cos\theta_{e\gamma}=-0.99$, the value of the integral increases 
by more than 3 orders of magnitude for $x_{min}=y_{min}\ge 0.95$.

On the other hand, such a change in the analysis is prone to 
induce background events, which the MEG cuts were designed for 
to avoid. The MEG proposal discusses as the two most important 
sources of background: (a) Prompt events from the standard model 
radiative decay $\mu \to e \nu {\bar\nu} \gamma$; and (b) accidental 
background from muon annihilation in flight. For the current 
experimental setup the accidental background is larger than the 
prompt background. Certainly, a better timing resolution of the 
experiment would be required to reduce this background. For the 
prompt background we estimate, using the formulas of \cite{Kuno:1999jp}, 
that for a total of $10^{13}$ muon events, one background event 
from the radiative decay will enter the analysis window for 
$x_{min}=y_{min} \simeq 0.96$ for the current cut on $\cos\theta_{e\gamma}$.

A further relaxation of the cuts can lead, in principle, to much 
larger values for ${\cal I}(x_{min},y_{min})$. However, the search 
for Br$(\mu\to e J\gamma)$ than necessarily is no longer background 
free. Since all the events from $\mu\to e J\gamma$ lie along the line 
of $\cos\theta_{e\gamma}$ defined by eq. (\ref{eq:ctheta}), whereas 
events from the SM radiative mode fill all of the $\cos\theta_{e\gamma}$
space, such a strategy might be advantageous, given a large enough 
data sample. 

Before closing this section, we mention that tau decays with Majoron 
emssion are less interesting phenomenologically for two reasons. 
First the existing experimental limits are much weaker for taus 
\cite{Baltrusaitis:1985fh} Br($\tau \to \mu J$) $\le 2.3$ $\%$  and 
Br($\tau \to e J$) $\le 0.73$ $\%$. And, second, although the coupling 
$\tau-\mu-J$ is larger than the coupling $\mu-e-J$ by a factor 
$m_{\tau}/m_{\mu}$, the total width of the tau is much larger than 
the width of the muon, thus the resulting theoretical predictions for 
tau branching ratios to Majorons are actually smaller than for the 
muon by a factor of approximately $10^4$.

\section{Conclusions}
\label{sec:cncl}

We have calculated branching ratios for exotic muon and tau 
decays involving Majorons in the final state. Branching ratios 
can be measurably large, if the scale of lepton number breaking 
is low. This result is independent of the absolute value of the 
neutrino mass. The lowest possible values of $v_R$ (at large 
values of $v_{\phi}$) are already explored by the existing limit 
on Br($\mu\to e J$). 

We have briefly discussed the status of experimental limits. It 
will not be an easy task to improve the current numbers in future 
experiments. While MEG \cite{meg} certainly has a high number of 
muon events in the detector, a search for Br$(\mu\to e J\gamma)$ 
instead of Br$(\mu\to e J)$ suffers from a small value of the 
available phase space integral, given current MEG cuts. An 
improvement will only be possible, if a dedicated search by the 
experimentalists is carried out. Nevertheless, we believe this 
is a worthwhile undertaking, since measuring a finite value for 
Br($\mu\to e J$) will establish that R-parity is broken in a 
region of SUSY parameter space complementary to that probed by 
accelerator searches.

\section*{Acknowledgments}

Work supported by Spanish grants FPA2008-00319/FPA and Accion Integrada
NO HA-2007-0090 (MEC). A.V. thanks the Generalitat Valenciana for 
financial support. W.P.~is supported by the DAAD, project number 
D/07/13468, and partially by the German Ministry of Education and 
Research (BMBF) under contract 05HT6WWA.

\bibliographystyle{h-physrev}

\begin{thebibliography}{10}

\bibitem{Chikashige:1980ui}
  Y.~Chikashige, R.~N.~Mohapatra and R.~D.~Peccei,
  Phys.\ Lett.\  B {\bf 98}, 265 (1981).

\bibitem{Gelmini:1980re}
  G.~B.~Gelmini and M.~Roncadelli,
  Phys.\ Lett.\  B {\bf 99}, 411 (1981).

\bibitem{Aulakh:1982yn}
  C.~S.~Aulakh and R.~N.~Mohapatra,
  Phys.\ Lett.\  B {\bf 119}, 136 (1982).

\bibitem{Amsler:2008zz}
  C.~Amsler {\it et al.}  [Particle Data Group],
  Phys.\ Lett.\  B {\bf 667}, 1 (2008).

\bibitem{Georgi:1981pg}
Majoron emission in neutrinoless 
double beta decay has been first discussed in: 
H.~M.~Georgi, S.~L.~Glashow and S.~Nussinov,
Nucl.\ Phys.\  B {\bf 193}, 297 (1981);
For reviews on double beta decay, see for example: 
M.~Doi, T.~Kotani and E.~Takasugi,
Prog.\ Theor.\ Phys.\ Suppl.\  {\bf 83}, 1 (1985);
W.~C.~Haxton and G.~J.~Stephenson,
Prog.\ Part.\ Nucl.\ Phys.\  {\bf 12} (1984) 409.

\bibitem{Arnold:2006sd}
  R.~Arnold {\it et al.}  [NEMO Collaboration],
  Nucl.\ Phys.\  A {\bf 765}, 483 (2006)
  [arXiv:hep-ex/0601021];
  For updated information on 
  the experiment see, for example:  
  L.~Simard,
  arXiv:0810.0533 [hep-ex].

\bibitem{Raffelt:1990yz}
  G.~G.~Raffelt,
  Phys.\ Rept.\  {\bf 198}, 1 (1990);
  G.~G.~Raffelt,
``Stars As Laboratories For Fundamental Physics: The Astrophysics Of
Neutrinos, Axions, And Other Weakly Interacting Particles,'' 
{\it  Chicago, USA: Univ. Pr. (1996) 664 p};
  G.~G.~Raffelt,
  Ann.\ Rev.\ Nucl.\ Part.\ Sci.\  {\bf 49}, 163 (1999)
  [arXiv:hep-ph/9903472].

\bibitem{Kachelriess:2000qc}
  M.~Kachelriess, R.~Tomas and J.~W.~F.~Valle,
  Phys.\ Rev.\  D {\bf 62}, 023004 (2000)
  [arXiv:hep-ph/0001039].


\bibitem{Hirsch:2006di}
M.~Hirsch and W.~Porod,
Phys.\ Rev.\  D {\bf 74}, 055003 (2006)
[arXiv:hep-ph/0606061].


\bibitem{Hirsch:2008ur}
  M.~Hirsch, A.~Vicente and W.~Porod,
  Phys.\ Rev.\  D {\bf 77}, 075005 (2008)
  [arXiv:0802.2896 [hep-ph]].

\bibitem{Masiero:1990uj}
  A.~Masiero and J.~W.~F.~Valle,
  Phys.\ Lett.\  B {\bf 251}, 273 (1990).

\bibitem{Romao:1991tp}
  J.~C.~Romao, N.~Rius and J.~W.~F.~Valle,
  Nucl.\ Phys.\  B {\bf 363}, 369 (1991).

\bibitem{meg}
Proposal to PSI:  
``MEG: Search for $\mu \to e \gamma$ down to $10^{-14}$ branching ratio'';
Documents and status at http://meg.web.psi.ch/. 
For a status report see, for example: 
  A.~Maki,
  AIP Conf.\ Proc.\  {\bf 981} (2008) 363.

\bibitem{Hirsch:2004rw}
M.~Hirsch, J.~C. Romao, J.~W.~F. Valle and A.~Villanova~del Moral,
\newblock Phys. Rev. {\bf D70}, 073012 (2004)

\bibitem{Hirsch:2005wd}
M.~Hirsch, J.~C. Romao, J.~W.~F. Valle and A.~Villanova~del Moral,
\newblock Phys. Rev. {\bf D73}, 055007 (2006)

\bibitem{Porod:2003um}
  W.~Porod,
  Comput.\ Phys.\ Commun.\  {\bf 153}, 275 (2003)

\bibitem{Maltoni:2004ei}
M.~Maltoni, T.~Schwetz, M.~A.~Tortola and J.~W.~F.~Valle,
New J.\ Phys.\  {\bf 6}, 122 (2004); 
An updated analysis with data included up to 
July 2008 has been published in 
T.~Schwetz, M.~Tortola and J.~W.~F.~Valle,
arXiv:0808.2016 [hep-ph].

\bibitem{Jodidio:1986mz}
  A.~Jodidio {\it et al.},
  Phys.\ Rev.\  D {\bf 34}, 1967 (1986)
  [Erratum-ibid.\  D {\bf 37}, 237 (1988)].

\bibitem{Picciotto:1987pp}
  C.~E.~Picciotto {\it et al.},
  Phys.\ Rev.\  D {\bf 37}, 1131 (1988).

\bibitem{Goldman:1987hy}
  J.~T.~Goldman {\it et al.},
  Phys.\ Rev.\  D {\bf 36}, 1543 (1987).

\bibitem{Kuno:1999jp}
  Y.~Kuno and Y.~Okada,
  Rev.\ Mod.\ Phys.\  {\bf 73}, 151 (2001)
  [arXiv:hep-ph/9909265].

\bibitem{Baltrusaitis:1985fh}
  R.~M.~Baltrusaitis {\it et al.}  [MARK-III Collaboration],
  Phys.\ Rev.\ Lett.\  {\bf 55}, 1842 (1985).




\end{thebibliography}

\end{document}